\documentclass[sigconf,table,screen]{acmart}

\acmConference[Accepted at ESEC/FSE 2022]{The 30th ACM Joint European Software Engineering Conference and Symposium on the Foundations of Software Engineering}{14 - 18 November, 2022}{Singapore}

\usepackage{xspace}
\usepackage{cleveref}
\usepackage{todonotes}

\usepackage[utf8]{inputenc}
\usepackage[english]{babel}
\usepackage{tabularx}
\usepackage{booktabs}
\usepackage{latexsym}
\usepackage{mathrsfs}
\usepackage{enumerate}
\usepackage{colortbl}
\usepackage{longtable}
\usepackage{tabu}
\usepackage{balance}
\usepackage{multirow}
\usepackage{epstopdf}
\usepackage{paralist}
\usepackage{adjustbox}
\usepackage{color, colortbl}
\usepackage[inline]{enumitem}
\usepackage{filecontents}
\usepackage{listings}
\usepackage{tcolorbox}
\usepackage{arydshln}
\usepackage{nccmath}
\usepackage{lipsum}
\usepackage{pdfcomment}  
\usepackage{url}
\usepackage{listings}
\usepackage{tcolorbox}
\usepackage{cleveref}[noabbrev]

\usepackage{adjustbox}
\usepackage[T1]{fontenc}
\usepackage{upquote}
\usepackage{graphicx}
\usepackage{hhline}
\usepackage{subcaption}
\usepackage{amsmath} 
\usepackage{soul} 


\setcopyright{none}
\renewcommand\footnotetextcopyrightpermission[1]{}
\copyrightyear{2022} 
\acmYear{2022} 
\setcopyright{acmlicensed}\acmConference[ESEC/FSE '22]{Proceedings of the 30th ACM Joint European Software Engineering Conference and Symposium on the Foundations of Software Engineering}{November 14--18, 2022}{Singapore, Singapore}
\acmBooktitle{Proceedings of the 30th ACM Joint European Software Engineering Conference and Symposium on the Foundations of Software Engineering (ESEC/FSE '22), November 14--18, 2022, Singapore, Singapore}
\acmPrice{15.00}
\acmDOI{10.1145/3540250.3549177}
\acmISBN{978-1-4503-9413-0/22/11}

\begin{document}

\title[The Impact of File Position on Code Review]{First Come First Served:\\ The Impact of File Position on Code Review}

\author{Enrico Fregnan, Larissa Braz}
\email{{fregnan,larissa}@ifi.uzh.ch}
\affiliation{%
  \institution{University of Zurich}
  \country{Switzerland}
  }

\author{Marco D'Ambros}
\email{marco.dambros@usi.ch}
\affiliation{%
	\institution{CodeLounge at Software Institute}
	\country{Università della Svizzera Italiana, Switzerland}
}

\author{G\"{u}l \c{C}alikli}
\email{handangul.calikli@glasgow.ac.uk}
\affiliation{%
	\institution{University of Glasgow}
	\country{Scotland}}

\author{Alberto Bacchelli}
\email{bacchelli@ifi.uzh.ch}
\affiliation{%
	\institution{University of Zurich}
	\country{Switzerland}
}

\renewcommand{\shortauthors}{}

\newcommand\etc{\emph{etc.}\xspace}
\newcommand\eg{\emph{e.g.},\xspace}
\newcommand\ie{\emph{i.e.},\xspace}

\newcommand{\numberPRs}{219,476\xspace}
\newcommand{\numberProjects}{138\xspace}

\newcommand{\expAccessed}{372\xspace}
\newcommand{\validParticipants}{106\xspace}
\newcommand{\participantsCC}{56\xspace}
\newcommand{\participantsMB}{50\xspace}

\newcommand{\participantsfoundCC}{52\xspace}
\newcommand{\participantsfoundMB}{50\xspace}

\newcommand{\toolName}{CRExperiment\xspace}

\newcommand\cornercase{Corner Case defect\xspace}
\newcommand\missingbreak{Missing Break defect\xspace}

\newcommand\tOne{\mbox{CC$_{f}$-MB$_{l}$}}
\newcommand\tTwo{\mbox{MB$_{f}$-CC$_{l}$}}

\newcommand\tenpow[1]{\ensuremath{{\times}10^{#1}}}

\definecolor{gray50}{gray}{.5}
\definecolor{gray40}{gray}{.6}
\definecolor{gray30}{gray}{.7}
\definecolor{gray20}{gray}{.8}
\definecolor{gray10}{gray}{.9}
\definecolor{gray05}{gray}{.95}

\newlength\Linewidth
\def\findlength{\setlength\Linewidth\linewidth
	\addtolength\Linewidth{-4\fboxrule}
	\addtolength\Linewidth{-3\fboxsep}
}

\newenvironment{rqbox}{\par\begingroup
	\setlength{\fboxsep}{5pt}\findlength
	\setbox0=\vbox\bgroup\noindent
	\hsize=0.95\linewidth
	\begin{minipage}{0.95\linewidth}\normalsize}
	{\end{minipage}\egroup
	\textcolor{gray20}{\fboxsep1.5pt\fbox
		{\fboxsep5pt\colorbox{gray05}{\normalcolor\box0}}}
	\endgroup\par\noindent
	\normalcolor\ignorespacesafterend}
\let\Examplebox\examplebox
\let\endExamplebox\endexamplebox

\newcommand{\rb}[1]{
	
	\begin{tcolorbox}[colback=gray!05,
		colframe=black,
		width=\columnwidth,
		arc=3mm, auto outer arc,
		boxrule=0.5pt,
		]
		#1
	\end{tcolorbox}
}

\newcounter{Finding}
\stepcounter{Finding}

\newcommand{\roundedbox}[1]{
	\rb{
		\noindent
		\textit{\textbf{Finding \theFinding}. #1}
	}
	\stepcounter{Finding}
}


\newboolean{showcomments}
\setboolean{showcomments}{true}
	
\ifthenelse{\boolean{showcomments}}
{
	\newcommand{\nb}[3]{
		{\colorbox{#2}{\bfseries\sffamily\scriptsize\textcolor{white}{#1}}}
		{\textcolor{#2}{\textsf\small$\blacktriangleright$\textit{#3}$\blacktriangleleft$}}}
	 \newcommand{\version}{\emph{\scriptsize$-$Id$-$}}
}{
	\newcommand{\nb}[3]{}
} 

\newcommand{\hide}[1]{}

\definecolor{purple}{HTML}{DADAEB}
\definecolor{blue1}{HTML}{e1effc}
\definecolor{babypink}{HTML}{ffedf8}
\definecolor{palevioletred}{HTML}{db7093}

\newcommand{\ab}[1]{\nb{Alberto}{palevioletred}{#1}}
\newcommand{\gul}[1]{\nb{Gül}{orange}{#1}}

\newcommand{\bluetext}[1]{\textcolor{black}{#1}}

\begin{abstract}

The most popular code review tools (\eg Gerrit and GitHub) present the files to review sorted in alphabetical order. Could this choice or, more generally, the relative position in which a file is presented bias the outcome of code reviews? 
We investigate this hypothesis by triangulating complementary evidence in a two-step study.

First, we observe developers' code review activity. We analyze the review comments pertaining to \numberPRs Pull Requests (PRs) from \numberProjects popular Java projects on GitHub. 
We found files shown earlier in a PR to receive more comments than files shown later, also when controlling for possible confounding factors: \eg the presence of discussion threads or the lines added in a file. 
Second, we measure the impact of file position on defect finding in code review. 
Recruiting \validParticipants participants, we conduct an online controlled experiment in which we measure participants' performance in detecting two unrelated defects seeded into two different files. Participants are assigned to one of two treatments in which the position of the defective files is switched. For one type of defect, participants are not affected by its file's position; for the other, they have 64\% lower odds to identify it when its file is last as opposed to first.
Overall, our findings provide evidence that the relative position in which files are presented has an impact on code reviews' outcome; we discuss these results and implications for tool design and code review.


\noindent
\textbf{Data and Materials:} \url{https://doi.org/10.5281/zenodo.6901285}

\end{abstract}


\maketitle

\newpage
\section{Introduction}

Code review is a popular software engineering practice where developers manually inspect the code written by a peer~\cite{Bacchelli2013, Macleod2017}. Code review aims to find defects~\cite{baum2016need}, improve software quality~\cite{ackerman1989software, Baum2017}, and transfer knowledge~\cite{Bacchelli2013, Sadowski2018}.
Over the years, code review has evolved from a formal strictly-regulated process~\cite{Fagan1976} into a less strict practice. Contemporary code reviewing is informal, asynchronous, change-based, and supported by tools~\cite{Sadowski2018, Rigby2013, Baum2016, baum2017choice}.


The tools used to conduct code reviews share many similarities~\cite{baum2016need}. In particular, the vast majority of tools (including the popular Gerrit~\cite{gerrit} and GitHub~\cite{GitHub}) present the changes to review as a list/sequence of diff hunks~\cite{hunk} grouped by the file they belong to. Tools sort these files alphabetically, therefore the changes to a file named \texttt{org/Controller.java} are always presented before those to a file named \texttt{org/Model.java}. Could this choice or, more generally, the relative position in which a file is presented influence the outcome of code review?

This hypothesis seems to be supported by at least two factors. First, most developers tend to start their reviews in the order presented by the review tool~\cite{Baum2017}. Second, code review is a cognitively demanding task~\cite{baum2019thesis} whose outcome might be influenced by cognitive factors~\cite{mohanani2018cognitive, spadini2020primers} also related to the position of the file. For example, developers may be influenced by \textit{attention decrement} (a decrease in attention when exposed to a list of elements~\cite{psyDictionary}) or may deplete their \emph{working memory capacity} (the memory for short-term storage during ongoing tasks~\cite{wilhelm2013working}) near the end of longer reviews. 

In this paper, we set to investigate this hypothesis. We do this by triangulating complementary evidence in a two-step study.

In the first step, we focus on the relation between file position and reviewers' activity. We collect and analyze \numberPRs Pull Requests (PRs) from \numberProjects GitHub open-source Java projects and investigate whether the position in which a file is presented in a PR is associated with the number of comments the file receives. In fact, the number of comments can be used to approximate reviewers' effectiveness and activity~\cite{Rigby2013}. We find a significant correlation between the file's position and the number of comments this file receives: Files shown later in a PR receive fewer comments than files shown earlier.

In the second step, we focus on the influence of file position on defect finding in code review. We design and conduct a controlled experiment with \validParticipants developers who have to review code in which we seeded two unrelated bugs (a \cornercase and a \missingbreak) into two different files. By creating two treatments that switch the position of the defective files (first \cornercase and last \missingbreak, or vice versa), we measure the influence of file position. While we see no effect for the \missingbreak, we find that developers have 64\% lower odds to identify the \cornercase when the file containing it is displayed last as opposed to first. To further confirm our findings, we look at the files displayed on the participants' screen during the review task. We detect a statistically significant difference between the time participants spend viewing the first file and the last one.

Overall, our findings suggest that the relative position in which files are presented in a code review has an impact on the code review's outcome. This result has important implications for code review practice and for tool design. For instance, code review tool designers may consider rethinking alphabetically ordering files in favor of a more principled approach (\eg showing first the most problematic files to review). Alternatively, code change authors may consider clearly signaling developers where to start their review, for example through the use of self-comments or in the change description, when they implemented more challenging parts.


\section{Background and Related Work}

In this section, we present relevant literature on cognitive aspects in modern code review. Then, we illustrate the psychological concepts that might constitute the underlying causes of the effect of file position on code review. Finally, we report possible competing arguments against the hypothesis that file position plays a role in reviewing.

\subsection{Cognitive Aspects in Code Review}

Code review is a collaborative process where human factors play a crucial role~\cite{Bacchelli2013}. Previous studies conducted at companies such as Microsoft~\cite{Bacchelli2013} and Google~\cite{Sadowski2018} revealed how code review could foster knowledge transfer among developers in a team and improve shared code ownership. However, code review is not only a collaborative process, but it is also a cognitively demanding task for the single reviewer. In particular, a vast amount of research focused on reducing developers' cognitive load to improve reviewers' performance~\cite{baum2016need, Baum2019, Gonccalvesexplicit}. For instance, \citeauthor{Baum2019}~\cite{Baum2019} conducted a controlled experiment to investigate how developers' code review performance relates to their cognitive load and working memory capacity. Their findings highlighted how working memory is associated with finding delocalized defects while only weakly associated with the ability to detect other defect types. 

Another group of studies addresses developers' cognitive biases that might affect code review outcome~\cite{spadini2020primers, thongtanunam2021reviewdynamics, chattopadhyay2020talefromthetrenches, huang2020biases}. \citeauthor{ huang2020biases} investigated how cognitive biases relate to code review process in a controlled experiment using medical imaging and eye-tracking. \citeauthor{chattopadhyay2020talefromthetrenches} synthesized helpful practices to reduce the effect of cognitive biases on software development activities, including code review. In their two-part field study, the authors identified the manifestation of various cognitive biases during code reviews (e.g., representativeness, availability, anchoring, confirmation bias). \citeauthor{spadini2020primers}~\cite{spadini2020primers} investigated the effect of \textit{priming} on reviewers' ability to detect bugs due to their proneness to availability bias. In the controlled experiment, a review comment was used to prime the participants to look for defects of the same type. The authors initially hypothesized that the participants would overlook defects of other types present in the code due to their proneness to availability bias. However, the experiment results show that the presence of a review comment identifying a bug does not prime reviewers towards looking for defects of similar type overlooking other defect types. On the contrary, such a review comment acts as a reminder for bugs developers usually do not look for during their daily practices. 

\subsection{Psychological Factors Related to Position}

Based on the findings of \citeauthor{Baum2017}~\cite{Baum2017}, which suggest that developers start a code review from the first file in the change-set, the following psychological factors seem plausible explanations behind the potential effect of files' position on code review. 


\smallskip
\noindent\textbf{Attention decrement.} Attention decrement is defined as ``\textit{the tendency for people to pay less attention to stimuli coming later in a sequential occurrence or presentation and thus to remember them less well}''~\cite{psyDictionary}. To give a practical example, a student given a list of terms to memorize is likely to have more difficulties committing to memory those at the end. 
\citeauthor{hendrick1973attention}~\cite{hendrick1973attention} identified attention decrement as the most feasible explanation of why people tend to remember the first information they see the most.

During code review, developers might be subjected to this phenomenon, slowly decreasing their attention the more the review continues. As the majority of reviewers begin their review from the first file in a change-set~\cite{Baum2017}, this decrease in attention is more likely to impact the last files in a review change-set. This would sustain our hypothesis that the position of the files in code review matters. 


\smallskip
\noindent
\textbf{Working memory.}  Working memory is defined as the part of human memory needed for short-term storage during ongoing cognitive processes~\cite{Baum2019, wilhelm2013working}. However, although the working memory capacity varies from individual to individual, its amount is still limited~\cite{cowan2010magical}.

In software engineering, the study of \citeauthor{bergersen2011programming} investigated the effect of working memory on programming performance~\cite{bergersen2011programming}. Their findings showed how working memory influences performance (albeit mediated by the developers' programming knowledge). \citeauthor{Baum2019} studied the effect of working memory on code review performance~\cite{Baum2019}. Their experiment showed a correlation between working memory and the reviewers' effectiveness in finding delocalized defects.
Reviewers might deplete their working memory looking at the first files in a review change-set, leading to an exhaustion of this resource that might negatively influence their code review performance on the last files they inspect.

\subsection{Related Competing Arguments}

Although previous evidence~\cite{Baum2017} suggests that developers start a code review from the first file in the change-set, some factors might affect the order in which developers review files during code review.
Given the findings of some eye-tracking studies (\eg~\cite{Abid2019developerreading,rodeghero2015eye-tracking,busjahn2015eye}), it seems plausible to think that some files in a code change might attract reviewers' attention more due to their content (\eg files that contain more call terms, control flow terms) regardless of their position. For instance, \citet{Abid2019developerreading} found that developers visit call terms more often than non-call terms and spend the longest time reading call terms. The next most visited and read locations were control flow terms and signatures. Furthermore, in an earlier study \citet{rodeghero2015eye-tracking} found that developers consider method signatures as the most important section of the code followed by call terms and the control flow terms. 
Moreover, token length and frequency in a source code might also influence developers' attention. The results \citeauthor{alMadi2021fromNovicetoExpert}~\cite{alMadi2021fromNovicetoExpert} obtained indicate that participants fixate longer low-frequency tokens and tokens containing more characters.  

Furthermore, in an eye-tracking study, \citet{busjahn2015eye} showed that people read source code in a more non-linear fashion than reading natural language.
However, the findings by \citeauthor{ busjahn2015eye} only explain how developers read code in a single code file: the authors showed the participants Java classes and pseudocodes ranging from a few lines of code to an entire screen full of text.
Therefore, whether developers navigate files in a code change in a non-linear fashion during code review and how this is related to file position need further investigation.   

In the presence of these arguments in the literature competing with the hypothesis that file location may influence code review outcome, our study aims to investigate this topic in-depth from complementary angles.

\section{Methodology}

In this study, we aim to understand whether the relative position in which a file is shown for review influences the code review process. 

\subsection{Research Questions}

We structured our investigation in two steps, which seek to collect complementary evidence. First, we investigate the relationship between file position and reviewers' activity. We do so by mining data from PRs belonging to a vast set of open source projects and answer the following research question:

\begin{center}
	\begin{rqbox}
		\begin{description}	    
			\item[\textbf{RQ$_1$.}] To what extent is the relative position of a file in a review associated with the amount of comments the file receives?
		\end{description} 
	\end{rqbox}
\end{center} 


Second, we focus on the influence of file position on defect finding in code review. We do so by performing an online controlled experiment with software developers who have to conduct a review of code and answer the following question:

\begin{center}
	\begin{rqbox}
		\begin{description}
			\item[\textbf{RQ$_2$.}] What is the effect of the relative position of a defective file in a review on bug detection?
		\end{description} 
	\end{rqbox}
\end{center}

\subsection{{RQ$_1$} - File Position and Review Activity}

\noindent\textbf{Subject Projects.} For our analysis, we select \numberProjects open-source projects from GitHub, focusing on Java projects with a star-count above 1,000. As previously reported~\cite{borges2016understanding, blincoe2016understanding}, the number of stars of a project can be effectively used as an indication of the projects' popularity and health. 
We focus on Java as (1) it is a widely popular programming language~\cite{tiobe} and (2) focusing on one programming language might reduce potential bias introduced by different review practices of projects based on other programming languages. Moreover, we investigate large projects to reduce potential bias caused by project-specific review policies or characteristics. 

\smallskip
\noindent\textbf{Data Collection.} Using PyGitHub\footnote{PyGitHub: \url{https://github.com/PyGithub/PyGithub}} a Python library to access the GitHub REST API, for each PR, we extract the position of each file and the number of comments it receives.
Moreover, we collect other factors that can influence the number of comments a file receives (\ie confounding factors). We consider: (1) The number of lines added and (2) deleted in a file because larger changes may require more comments; 
(3) whether the file is a test, because these tend to receive less comments~\cite{spadini2018testing} and to be ordered last alphabetically; and
(4) the number of commenters, because more participants in the review of a PR might lead to more comments.

\smallskip
\noindent\textbf{Data Filtering.} As the presence of a bot might introduce bias in our analysis, we exclude bot comments from our analysis. To do so, we flag all users containing the word ``bot'' as bots and create a list of commonly used bots in GitHub (extracted from relevant gray literature~\cite{wesselOnline, twilio}) and remove them.
The presence of discussion threads (i.e., a series of comments where the reviewers and the author discuss solutions or improvements to the code) might act as a confounding factor in our analysis: A discussion thread can facilitate developers’ engagement in adding comments to the thread regardless of the file’s position. Therefore, we focus our investigation only on the first comment of a thread, disregarding subsequent ones.
%
Moreover, the number of files in a Pull Request might influence the comments’ distribution. Therefore, we group the PRs according to the number of files they contain to conduct initial analyses (\eg see \Cref{fig:prsummary5files} for PRs with five files).

\smallskip
\noindent\textbf{Data Analysis.} To analyze the impact of the aforementioned factors on the number of comments in a file (dependent variable), we build a \emph{Hurdle model}~\cite{feng2021comparison}: A statistical model that specifies two processes, one for zero counts and another for positive counts. We choose this model as it handles excess zeros in the dependent variables, in fact, a vast number of files in the collected PRs receive zero comments. To model the positive counts, we employ a \emph{negative binomial} distribution. Furthermore, we check the multi-collinearity across the variables in the model computing their Variance Inflation Factor (VIF) and removing those with a VIF higher than five.

\subsection{{RQ$_2$} - File Position and Defect Finding}\label{sec:met:rq2}

Our controlled experiment is organized as an online experiment in which participants are asked to complete a review of a code change involving five files. Among these files, we seed two unrelated software defects\footnote{The participants are not informed about the presence of these defects.} into two different files. We randomly assign participants to one of the following two treatments: (1) one file with a bug is presented first and the other file with the other bug is presented last (\ie fifth), or (2) the order of the files (hence the bugs) is reversed (\ie the fifth file is now first and vice versa).
We measure the specific bugs detected by each participant, as well as how long each file is visible on the participants' screen.

In the following, we provide more details on the \emph{experimental objects} (\ie code change to review and seeded defects) and the \emph{experimental design} (\ie online platform and experiment flow). We conclude by describing the pilots we conducted, how we recruited participants, and how we analyzed the collected experimental data.

\newpage
\noindent\textbf{3.3.1 Experimental Objects}

\smallskip
\noindent\textbf{Code change.}
For the review task, we create a code change that satisfies the following requirements: 
(1) not belonging to any existing project (to avoid introducing bias caused by participants' previous knowledge of the review change-set); 
(2) written in Java, one of the most popular programming languages~\cite{tiobe}; 
(3) self-contained (to minimize the previous knowledge participants need to understand the change fully); 
and (4) close to an actual code review scenario (to increase realism).

We create a code change spanning five files as a trade-off between small change-sets (\eg three files) and more complex reviews. For smaller code changes (\eg two files), we expect a smaller effect of a file's position. In contrast, even though larger code changes may present a stronger effect, more complex reviews require a longer review time, thus likely leading to more participants abandoning the experiment.

All files have similar sizes (ranging from 44 to 63 lines). In particular, we ensure that the two files containing the defects are similarly sized (46 lines in the \cornercase file and 44 lines in the \missingbreak) to minimize the potential bias introduced by the file length.  

We devise the code change to minimize links (\eg method calls) between non-adjacent files. In particular, we checked that no connection existed between the two defective files (\ie first and last). Our goal is to increase the control on how participants navigate the change, thus making the file position effect clearer, if any.

Finally, to re-position the files for the two different treatments, we rename them. This ensures that the reviewers are not influenced by a tool behavior that is unexpectedly different than what they are used to in common review platforms (where files are displayed in alphabetical order).

\begin{figure}[h]
	\centering
	\includegraphics[width=\columnwidth]{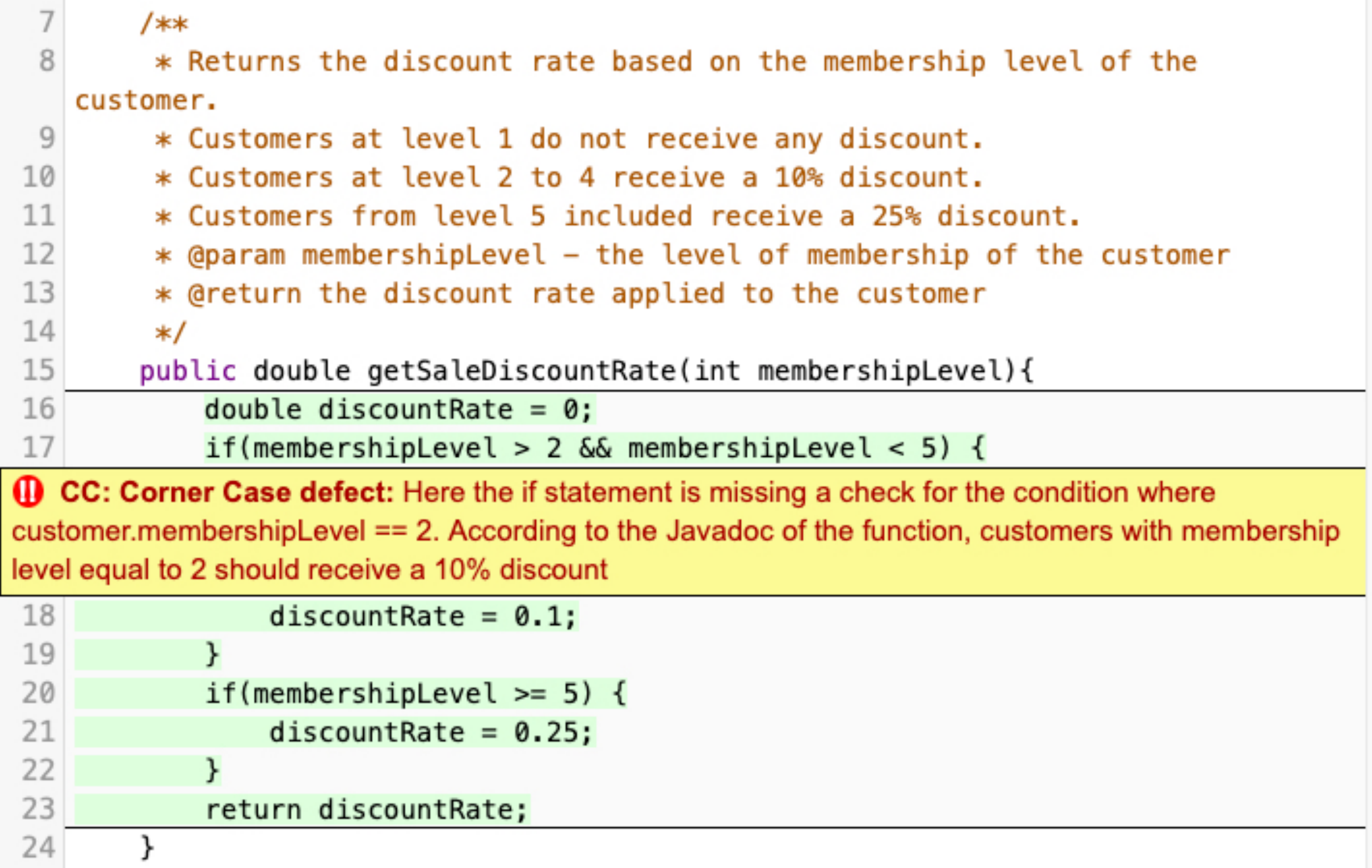}
	\caption{\cornercase (CC) used in our experiment.} 
	\label{fig:cornercase}
\end{figure}

\begin{figure}[h]
	\centering
	\includegraphics[width=\columnwidth]{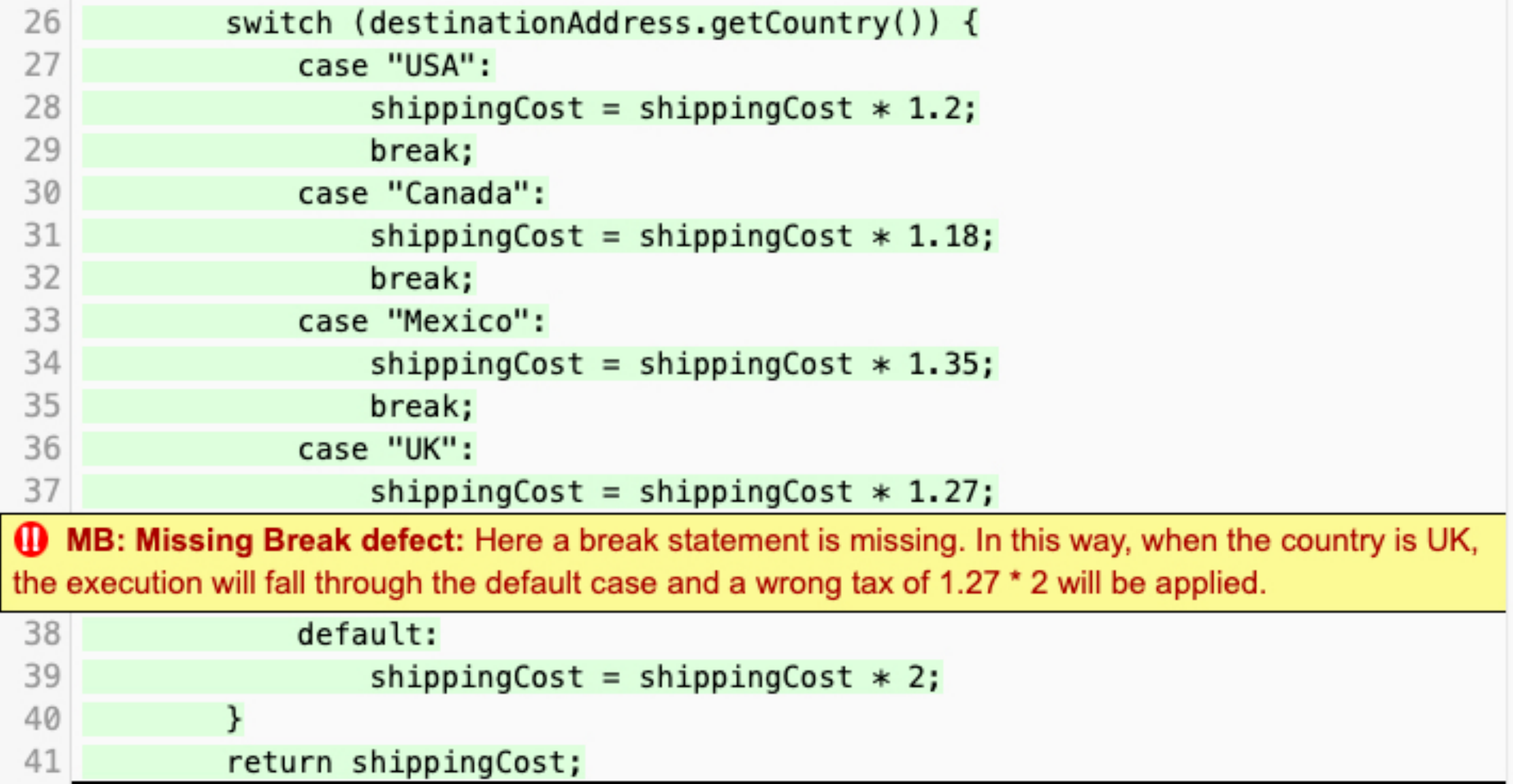}
	\caption{\missingbreak (MB) used in our experiment.} 
	\label{fig:missingbreak}
\end{figure}

\smallskip
\noindent\textbf{Seeded Defects.}
In our experiment, we investigate if the given treatment influences participants' ability to find bugs. 
In the review task, we seed the following two unrelated functional defects:

\begin{description}[leftmargin=0.4cm]
 \item[- Corner Case (CC):] A corner case condition in an \texttt{if} statement is left unchecked, thus not respecting the documentation (\Cref{fig:cornercase}). We select this bug because it represents an issue developers typically check for~\cite{spadini2020primers, braz2021don} and frequently exists in practice~\cite{jeng1994simplified}.
 
 \item[- Missing Break (MB):] A missing \texttt{break} statement in a \texttt{switch} construct makes the execution incorrectly fall through the default case (\Cref{fig:missingbreak}). We select this defect because it is reported as a common Java mistake by relevant gray literature~\cite{switch1, 5commonjavamistakes}, also resonating with the infamous Apple \textit{goto fail}~\cite{bland2014finding}.
 
\end{description}

\noindent
We use two defects 
to make the review task as realistic as possible by preventing participants from focusing exclusively on one bug.

\medskip
\smallskip
\noindent\textbf{3.3.2 Experimental Design}

\smallskip
\noindent\textbf{Experiment Platform.}
 We implement our experiment as an online platform. To this aim, we employ a publicly available tool, \toolName~\cite{crexperiment}, already successfully used in previous code review experiments~\cite{braz2021don, spadini2020primers}. \toolName allows participants to review code displaying a review change-set in two-pane diffs, as done in popular code review tools: \eg Gerrit~\cite{gerrit} or GitHub~\cite{GitHub}. Moreover, the User Interface (UI) of \toolName adopts similar design choices (\eg the color scheme) as the one of, for instance, Gerrit. This mitigates possible bias with the users not being familiar with the experiment platform. 

\toolName also logs participants' interactions (\eg scrolling) and the time participants spend in each phase of the study (\eg in the review tasks). 
To collect information on which files in the change-set the participants focus on during code review, we extended the base experiment platform to record the files visible on the participants' screen during the review. 
For each file, the experiment tool records if this was on the participants' screen (displayed), partially displayed, or not displayed at a given time.
We store all the collected data anonymously.




\begin{figure*}[h]
	\centering
	\includegraphics[width=\textwidth]{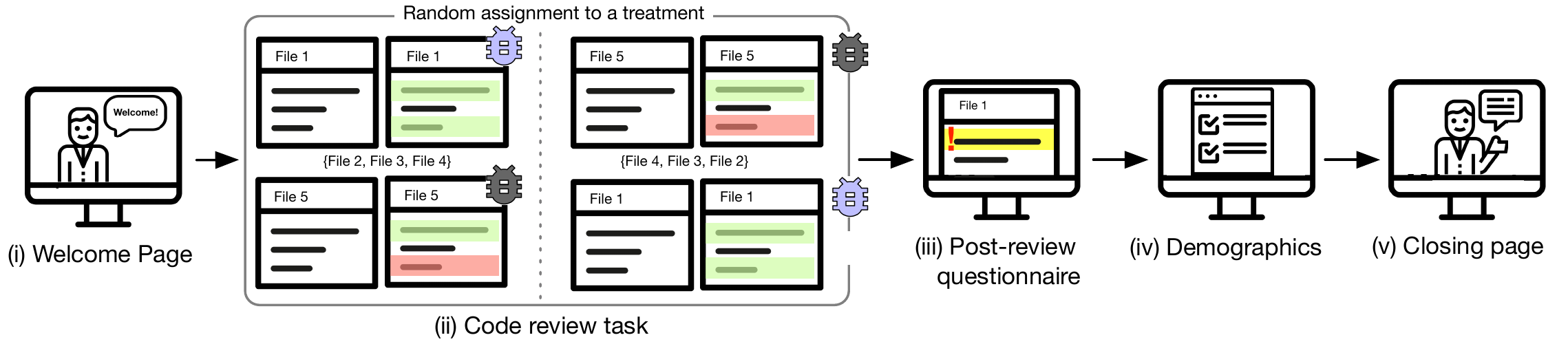}
	\caption{Design and flow of the online experiment.} 
	\label{fig:flow}
\end{figure*}

\smallskip
\noindent\textbf{Experiment flow.}
We organize the flow of our experiment as depicted in \Cref{fig:flow} and as described in the following:

\begin{description}[leftmargin=0.3cm]
 \item[(i) Welcome page.] On the experiment's landing page, we provide participants with information on the type of study and our data handling policy. Moreover, we ask for their consent to collect and use their data before proceeding in the experiment.
 
  \item[(ii) Code review task.] Participants are asked to perform a code review task. Before starting, we instruct participants to perform a code review as they would typically do in their normal practice.
  Participants are randomly assigned to one of two possible \textbf{treatments}:
  
  \begin{itemize}
    \item \textbf{\tOne} - The file containing the \cornercase is shown $f$irst in the code change, while the one containing the \missingbreak is shown $l$ast.
    \item \textbf{\tTwo} - The file containing the \missingbreak is shown $f$irst, while the one with the \cornercase is shown $l$ast.
  \end{itemize}
  
  To avoid bias, we do not inform participants about the functional defects, the treatments, and the recorded metrics before the review task. At the end of the task, we ask participants whether and for how long they were interrupted during the review.

  \item[(iii) Post-review questionnaire.]
After participants declare their review task is completed, they are shown the code change again with the location and explanation of the defects (\Cref{fig:cornercase} and \Cref{fig:missingbreak}). We ask participants whether they found each of the two defects and the cause that, in their opinion, influenced their result. In a subsequent page (as not to influence their previous answers), we ask participants if they think the position of the file containing the bugs might have impacted their results. 

  \item[(iv) Demographics.] Afterward, we collect information about participants' gender, education, occupation, and experience with Java and code review. We analyze this information (1) to guarantee that the groups of participants assigned to different treatments are homogeneous, (2) as confounding factors in the analysis of the experiment results, and (3) to describe the population represented in our results.
  
  \item[(v) Closing page.] On the final page of the experiment, we disclose the goal of our study and ask participants for any final remark. We also collect participants' consent to share their answers in an anonymized, yet publicly available, research dataset.

\end{description}

\medskip
\noindent\textbf{3.3.3 Pilot Runs}
	
\noindent Before publicly releasing the online experiment, we conduct a pilot study to (1) verify the absence of technical issues with experiment settings (\eg the online platform) and instructions, (2) check the goodness of the devised review task and seeded defects, and (3) improve the experiment based on the participants' feedback. 

We conduct four iterations of our pilot study with a total amount of 15 participants. After each iteration, the authors discuss the results and feedback obtained and improve the experiment design accordingly. After the last iteration,  the experiment is deemed ready to be released since the participants detected no significant issues.

We excluded the data gathered from the participants in the pilot study from the analysis of the experiment results. 

\medskip
\noindent\textbf{3.3.4 Data Analysis}

\noindent To analyze the data collected in our experiment, we perform a \textit{Chi-square} test to assess the correlation between files position and participants' detection of the defects. We evaluate the strength of the correlation using the \textit{phi-coefficient}.
Then, we build two logistic regression models considering as dependent variables (1) \textit{CCFound} (whether the participants found the \cornercase), and (2) \textit{MBFound} (whether the participants found the \missingbreak), respectively.

To identify if a participant found a defect, we follow the guidelines of previous studies~\cite{Baum2019, Gonccalvesexplicit, gonccalves2022explicit}. 
We manually inspect the remarks left by the participants and classify a defect as found when a remark (1) is placed in proximity of the defect and (2) clearly describes the defect in the code. The first author performs this inspection, and a second author checks later the first author's inspection results.

To build our models, we follow the following procedure to ensure the goodness of our results: (1) We use the VARCLUS procedure to identify and remove variables from the model with Spearman's correlation higher than 0.5. (2) We compute the Variance Inflation Factors (VIF) among the variables to ensure none had a VIF value above 7. If a variable with a higher VIF is detected, we remove it from the model. (3) Finally, we build the models adding each variable one by one and checking that the coefficient remained stable, showing little interference among the considered variables. 

We aim to evaluate if the position of the defect (in the first or last file) impacts participants' ability to find it. For this reason, we include the variable \textit{position} among the independent variables in our model. 
We also consider the potential effect of other confounding factors, such as the participants' experience with Java and code review. Furthermore, we compute the time participants spend on the review task and include this variable in the model. \Cref{tab:logVariables} presents all the variables included in our logistic regression models.

\begin{table}[ht]
	\centering 
	\caption{Variables used in the logistic regression models.} \label{tab:logVariables}
	\begin{tabular}{l|p{6cm}}
		 \multicolumn{1}{c}{\cellcolor[HTML]{C0C0C0}\textbf{Metric}} &  \multicolumn{1}{c}{\cellcolor[HTML]{C0C0C0}\textbf{Description}} \\
		 \multicolumn{2}{c}{\cellcolor[HTML]{E8E2E1}\textit{Dependent variables}}  \\
		CCFound & The participant found the \cornercase. \\
		MBFound & The participant found the \missingbreak. \\
		
		 \multicolumn{2}{c}{\cellcolor[HTML]{E8E2E1}\textit{Independent variables}}  \\
		Position & The relative position of the bug (first or last) in the code change to review.\\ 
	
		 \multicolumn{2}{c}{\cellcolor[HTML]{E8E2E1}\textit{Control variables}}  \\
		CRDuration & The time spent on the code review task.\\
		DevExp &  The participant's development experience in a professional setting.\\
		JavaExp & The participant's experience with Java.\\
		CRExp & The participant's experience in code review.\\
		OftenProg & How often the participant programmed in the three months prior to the experiment.\\
		OftenCR & How often the participant reviewed code in the three months prior to the experiment.\\
		Interruptions & How often and for how long the participant was interrupted during the review task.\\
		\hline
	\end{tabular}
\end{table}

\smallskip
\noindent
\textbf{3.3.5 Number of required participants}.

\noindent We performed an \emph{a priori} power analysis to compute the minimum sample size needed for our experiment. We used the G-Power software~\cite{faul2007g} and employed a \textit{two-tail} test (with a manual distribution) having $odds~ration = 1.5$, $\alpha = 0.05$, $Power = 0.8$, and $R^{2} = 0.3$. The results of this analysis showed our experiment needed a minimum of 92 valid participants.

	\smallskip
\noindent\textbf{3.3.6 Participants' recruitment}

\noindent We disseminated an invite to conduct the online experiment through the authors' professional network and social media accounts (\eg LinkedIn), as well as on developers' web forums and communication channels (\eg Reddit).
To prevent any bias in the results, we do not disclose the experiment's aim to participants in the invite. To encourage developers' participation, we commit to donate 5 USD to a charity for each valid participant in the experiment.

\section{Results}

We present the results of our investigation by research question.

\subsection{{RQ$_1$} -  File Position and Review Activity}

\begin{figure}[t]
	\centering
	\includegraphics[width=\columnwidth]{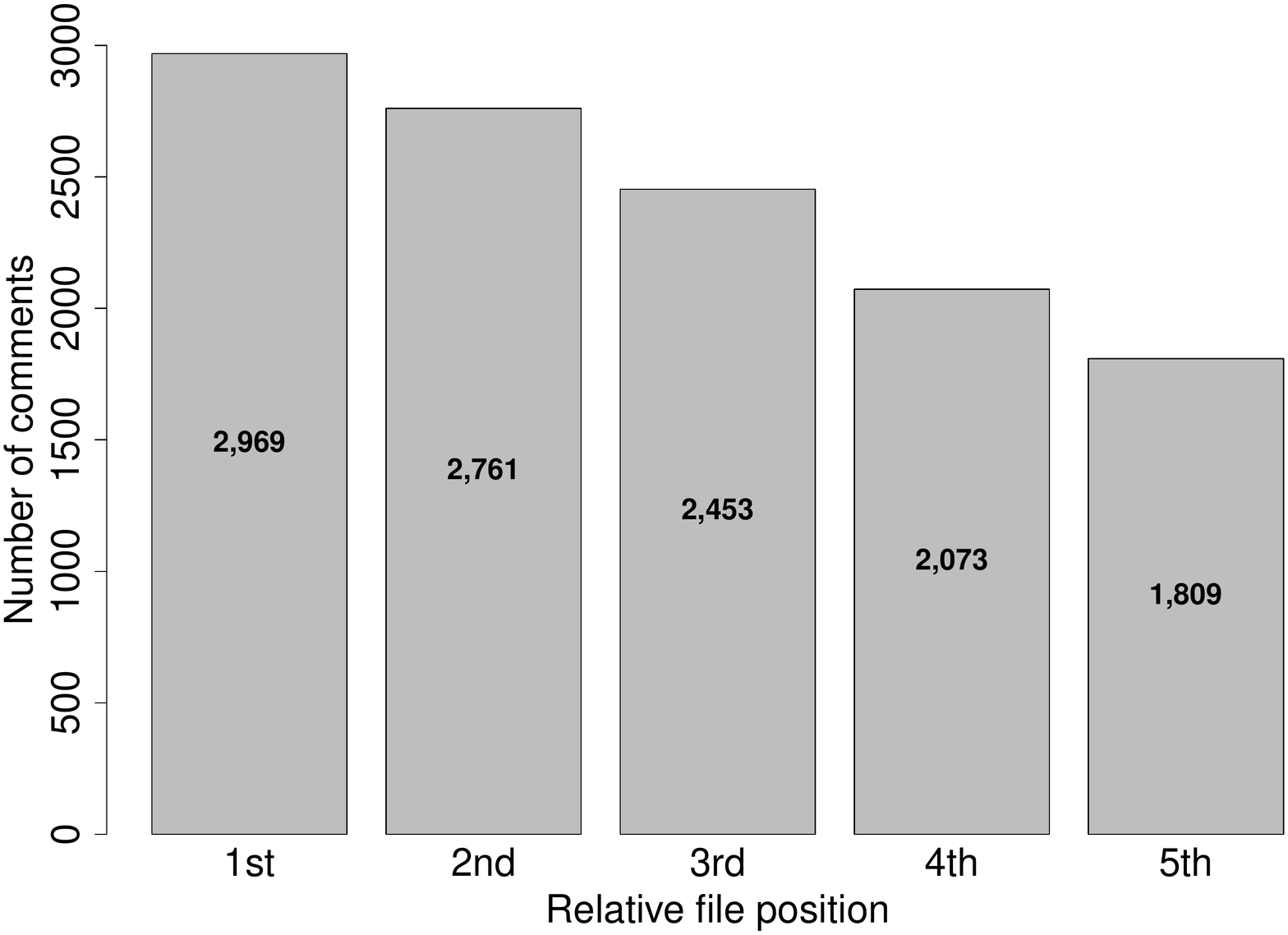}
	\caption{Distribution of comments in a PR with five files.}
	\label{fig:prsummary5files}
\end{figure}

Our first research question seeks to investigate the relationship between relative file position and reviewers' activity, focusing on the comments left during code review.
\bluetext{We analyze a total of 219,476 pull requests pertaining to 138 popular Java projects on GitHub. Among these, 26,685 pull requests received at least one review comment (which we computed excluding comments left by bots).}


As the vast majority of PRs (81.06\%) contain between one and ten files, we focused our investigation on this subset of PRs. We excluded from our analysis PRs containing only one file as the effect of the position in such cases is not relevant.

Code review on GitHub is an iterative process. When a developer uploads a commit, this is reviewed by fellow developers who may ask the commit author to perform some changes. These changes are addressed in a subsequent commit and the process continues iteratively until the code is accepted. To both consider and exclude the effect of the review process itself, we analyzed the data concerning two different moments in the history of pull requests:

\begin{description}[leftmargin=0.4cm]
 \item[- Pull Request summary]: We consider the code change as it appears at the end of the review process and all the comments that the code has received during the entire process.
 
 \item[- First commit]: We consider only the code as initially submitted in the pull request and the comments it receives. Thus, we exclude changes and comments induced by the review process.

\end{description}

\medskip
\noindent\textbf{4.1.1 Pull Request summary}
	
\noindent
Our initial investigation showed that the number of comments on the files in a PR is higher in the first files and progressively decreases towards the last files. For instance, \Cref{fig:prsummary5files} shows the distribution of the comments in PRs with five files. A similar pattern was observed also for all the other groups of PRs, \eg with six files. We found similar results also when we restrict our investigation to the PRs that contain only Java files.\footnote{Our online appendix~\cite{firstRepPackage} provides graphs for all the cases we mention.}

To confirm and quantify the correlation between the relative position of a file in a pull request and the number of comments it receives, we performed a \textit{Spearman correlation test} (a non-parametric test that measures the strength of the association between ordinal or continuous variables~\cite{spearmanCorr}). We obtained a p-value of $<2.2\cdot10^{-16}$ and a $\rho = -0.203$, thus confirming a statistically significant negative relation between these two variables.

\begin{table}[t]
	\centering 
	\caption{Hurdle models for PR summary.} \label{tab:hurdlePRSummary}
	{\rowcolors{3}{white}{gray05}
	\begin{tabular}{lrl|rl}
		& \multicolumn{2}{c|}{\cellcolor[HTML]{C0C0C0}\textbf{Positive count}} &  \multicolumn{2}{c}{\cellcolor[HTML]{C0C0C0}\textbf{Zero count}} \\
		&  \multicolumn{1}{c}{\cellcolor[HTML]{E8E2E1}\textbf{Estim.}} &  
		 \multicolumn{1}{c|}{\cellcolor[HTML]{E8E2E1}\textbf{Sig.}} &
		 \multicolumn{1}{c}{\cellcolor[HTML]{E8E2E1}\textbf{Estim.}} & 
		 \multicolumn{1}{c}{\cellcolor[HTML]{E8E2E1}\textbf{Sig.}}	\\
		\hline
		Intercept & -8.982 
		& ** & -3.571 
		&  ***  \\
		Position & -0.187 
		& *** & -0.343 
		& ***  \\
		Lines added & 0.012 
		&  *** & 0.007 
		 & ***  \\
		Lines deleted & 0.003 
		& *** & -0.001
		& *** \\
		isTest & -0.433 
		& *** & -0.707 
		& *** \\
		N. commenters & 0.485 
		 & *** & 1.142 
		 & *** \\
		\hline
		\multicolumn{5}{c}{Sig. codes: *** p < 0.001; ** p < 0.01; * p < 0.05} \\
	\end{tabular}
	}
\end{table}

Subsequently, we performed regression modeling to verify this association also when controlling for the other factors that may influence the number of review comments on a file. 

\Cref{tab:hurdlePRSummary} shows the Hurdle model predicting the number of comments in a file. We included in the model PRs containing only Java files to remove potential bias introduced by multiple file formats. Moreover, we excluded PRs having less than three files, since in these cases the effect of the file position is likely to be negligible. To remove outliers, we limited our analysis to files that have fewer than 1,000 lines added. Finally, we normalized each file's position over the number of files in the PR to allow for comparison among PRs with a different number of files. 
The model is summarized in \Cref{tab:hurdlePRSummary} and shows how the position of a file is negatively correlated with the number of comments it receives, confirming how the last files in a review change-set tend to receive less review comments from developers, also when controlling for other factors. 

\bluetext{\Cref{tab:hurdlePRSummary} also shows that the other independent variables included in the model are statistically significant predictors of the number of comments in a file. The sign of the estimates show that the direction of the relationship follows the expected direction. For instance, a larger number of changed lines (added or deleted) is related to more reviewers' comments. On the contrary, the expected number of comments decreases when the file is a test (this result is inline with previous research's findings~\cite{spadini2018testing}). Finally, the number of commenters is positively correlated with the number of comments. }

\medskip
\noindent\textbf{4.1.2 First commit}
	
\noindent
By restricting our analysis to only the first commits of our set of Pull Requests, we noticed a similar pattern as the one reported for the PR summary. The first files in a commit receive more comments compared to the last ones. 
We analyzed commits with number of files ranging from two to ten, noticing similar patterns to the PR summary (available in the replication package~\cite{firstRepPackage}).

As for the PR summary, we corroborated our findings by restricting our analysis to commits containing only Java files to mitigate potential bias caused by multiple types of files. 
This verification, however, did not reveal any significant difference from the general case confirming our observations that a higher number of comments concentrates on the first files in a commit.

We performed a \textit{Spearman correlation test} to verify the presence of a correlation between the number of review comments a file receives and the file's position. The test achieved a p-value of $<2.2\cdot10^{-16}$ and a $\rho = -0.152$, showing how the two variables are correlated (although this correlation is weaker).

To build the Hurdle model in this scenario, we followed the same steps as the one reported for the Pull request summary. Our results (reported in \Cref{tab:hurdleFirstCommit}) confirm how later files in a review change-set tend to receive a lower number of comments. 
\bluetext{Concerning the other independent variables included in our model, we achieved results similar to the ones presented for the PR summary (Section 4.1.1).}

\begin{table}[t]
	\centering 
	\caption{Hurdle models for PR's first commit.} \label{tab:hurdleFirstCommit}
	{\rowcolors{3}{white}{gray05}
	\begin{tabular}{lrl|rl}
		& \multicolumn{2}{c|}{\cellcolor[HTML]{C0C0C0}\textbf{Positive count}} &  \multicolumn{2}{c}{\cellcolor[HTML]{C0C0C0}\textbf{Zero count}} \\
		&  \multicolumn{1}{c}{\cellcolor[HTML]{E8E2E1}\textbf{Estim.}} &  
		\multicolumn{1}{c|}{\cellcolor[HTML]{E8E2E1}\textbf{Sig.}} &
		\multicolumn{1}{c}{\cellcolor[HTML]{E8E2E1}\textbf{Estim.}} & 
		\multicolumn{1}{c}{\cellcolor[HTML]{E8E2E1}\textbf{Sig.}}	\\
		\hline
		Intercept & -10.34 
		& . & -4.200 
		&  ***  \\
		Position & -0.183 
		& ** & -0.221 
		& ***  \\
		Lines added & 0.019 
		&  *** & 0.005 
		& ***  \\
		Lines deleted & 0.004 
		& *** & -$4.15\cdot10^{-4}$ 
		& * \\
		isTest & -0.353 
		& *** & -0.667 
		& *** \\
		N. commenters & 0.607 
		& *** & 1.685 
		& *** \\
		\hline
		\multicolumn{5}{c}{Sig. codes: *** p < 0.001;  ** p < 0.01;  * p < 0.05;  .  p < 0.1} \\
	\end{tabular}
	}
\end{table}

\smallskip
\roundedbox{The number of comments across the files in the analyzed pull requests is associated with the relative position of each file. Specifically, the later a file is presented in a pull request, the fewer the comments it receives.}


\subsection{{RQ$_2$} - File Position and Defect Finding}

Encouraged by the findings for {RQ$_1$}, we devised a controlled experiment to test this hypothesis with complementary evidence. Our second research question seeks to investigate the effect of files' position on developers' review effectiveness. 

A total of \expAccessed participants accessed the online experiment. Of these, we considered only the participants who completed all the steps. Furthermore, we removed participants who left no remarks and spent less than ten minutes doing the review, leaving us with  \validParticipants participants. 

\begin{figure*}[t]
		\centering
		\begin{subfigure}{.5\textwidth}
			\centering
			\includegraphics[width=\textwidth]{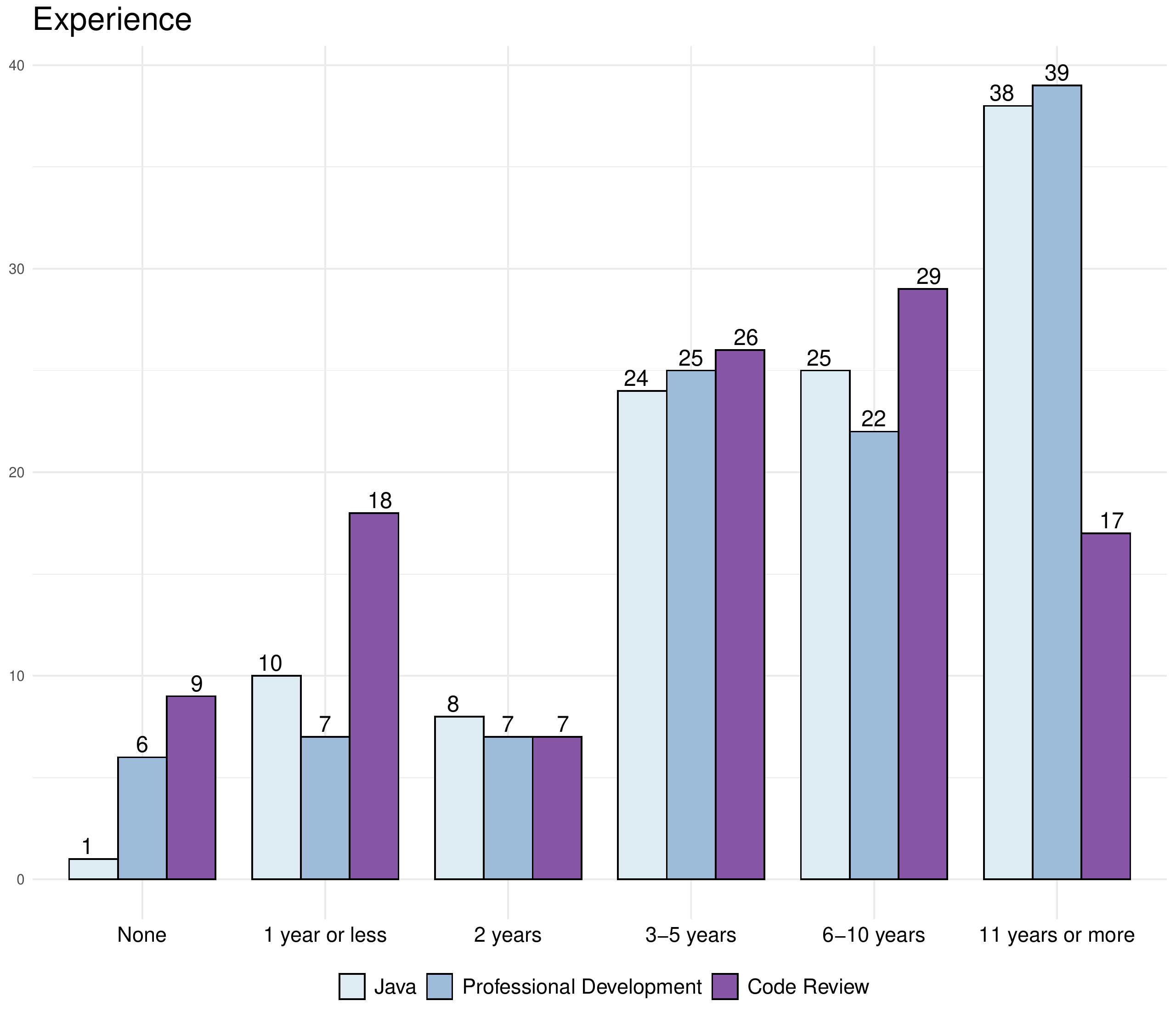}
			\label{fig:parsub1}
		\end{subfigure}%
		\begin{subfigure}{.5\textwidth}
			\centering
			\includegraphics[width=\textwidth]{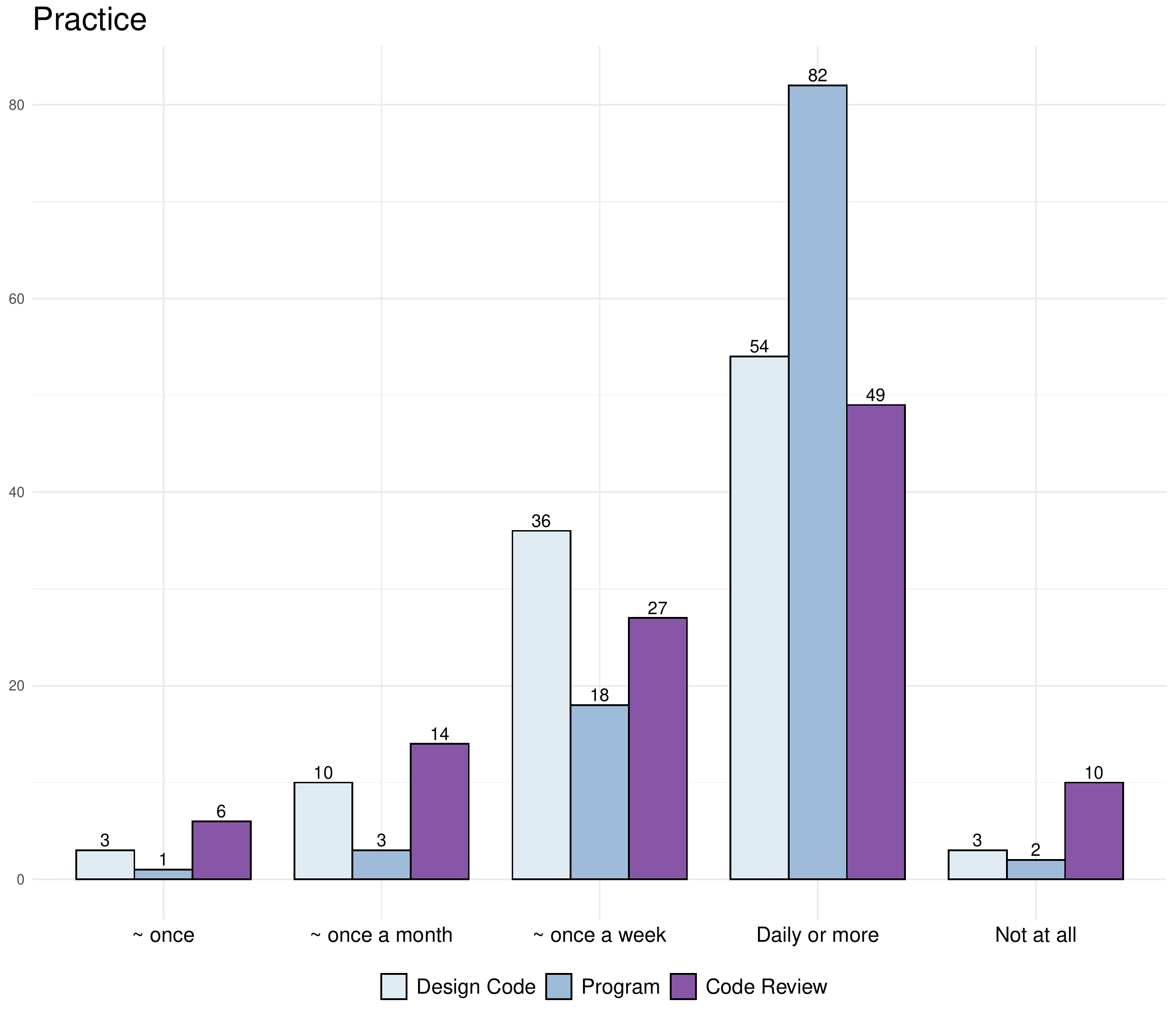}
			\label{fig:parsub2}
		\end{subfigure}
		\caption{Participants' experience and practice.}
		\label{fig:participantsDemo}
\end{figure*}

The vast majority of the participants ($84.9\%$) have at least a B.Sc. degree (mostly in Computer Science). Overall, 72 participants reported to be software developers. Moreover, 84 participants self-described as male, four as female, two as non-binary, and 16 preferred not to disclose. \Cref{fig:participantsDemo} reports participants' experience and practice levels.

In total, \participantsCC participants were in treatment \textbf{\tOne}, while \participantsMB were in treatment \textbf{\tTwo}. We compared the experience and practices (\eg Java experience) of the participants assigned to the two groups and found no statistically significant difference.

\medskip
\noindent\textbf{4.2.1 Defect Finding}
	
\noindent Overall, \participantsfoundCC participants found the \cornercase, while \participantsfoundMB detected the \missingbreak. 
\Cref{tab:bugsFoundTotal} reports the number of participants who found no defect, only one defect, or both defects in the review task, by treatment.
\Cref{tab:bugsFound} reports how many participants identified each type of bug, by treatment.

\begin{table}[ht]
	\centering 
	\caption{Participants who found (1) no defects, (2) only the \cornercase (CC), (3) only the \missingbreak (MB), and (4) both defects.} \label{tab:bugsFoundTotal}
	{\rowcolors{2}{white}{gray05}
	\begin{tabular}{rcrrc|r}
		 & 	\cellcolor[HTML]{C0C0C0}\textbf{No defect} & 	\cellcolor[HTML]{C0C0C0}\textbf{CC} 
		 & 	\cellcolor[HTML]{C0C0C0}\textbf{MB} & 	\cellcolor[HTML]{C0C0C0}\textbf{CC \& MB} & \cellcolor[HTML]{C0C0C0}\textbf{Total}\\
		\hline
		\textbf{\tOne} & 13 &  19 & 9 & 15 & 56  \\\hline
		\textbf{\tTwo} &17 & 7  & 15 & 11 & 50 \\ \hline
		\textbf{Total} & 30 & 26 & 24 & 26 & 106 
	\end{tabular}
	}
\end{table}
\begin{table}[ht]
	\centering 
	\caption{Participants who found/not found each bug divided by treatment: \cornercase (CC) first or \missingbreak (MB) first.} \label{tab:bugsFound}
	\begin{tabular}{lcccc}
		& \multicolumn{2}{c|}{\cellcolor[HTML]{C0C0C0}\textbf{Corner Case}} & \multicolumn{2}{c}{\cellcolor[HTML]{C0C0C0}\textbf{Missing Break}} \\
		&  \multicolumn{1}{c}{\cellcolor[HTML]{E8E2E1}\textbf{Found}} & \multicolumn{1}{c|}{\cellcolor[HTML]{E8E2E1}\textbf{Not found}} 
			& \multicolumn{1}{c}{\cellcolor[HTML]{E8E2E1}\textbf{Found}} & \multicolumn{1}{c}{\cellcolor[HTML]{E8E2E1}\textbf{Not found}} \\
		\hline
		\textbf{\tOne} & 34 & \multicolumn{1}{c|}{22} & 24& 32\\\hline
		\rowcolor{gray05}
		\textbf{\tTwo} &18 & \multicolumn{1}{c|}{32} & 26 & 24 \\ \hline
		& \multicolumn{2}{r}{p-value: \textbf{0.011}} & \multicolumn{2}{r}{p-value: 0.346} \\
		& \multicolumn{2}{r}{phi-coefficient: 0.247} & \multicolumn{2}{r}{phi-coefficient: 0.091} \\
		& \multicolumn{2}{r}{odds ratio: 2.75} & \multicolumn{2}{r}{odds ratio: 1.44} \\ 
	\end{tabular}
\end{table}

These results show how participants found more the first bug shown in the review task compared to the bug shown last. The \cornercase was found 34 times when the file containing it was shown first and only 18 when the file was shown last. A similar, albeit weaker, trend seems to appear for the \missingbreak.



To verify the presence of a relation between the position of a bug and its detection, we performed a \textit{Chi-square} test. In the case of the \cornercase, this test achieved ${\chi}^2 (df=1, N =106)=6.45)$, p-value = 0.011, rejecting the null hypothesis that no relationship exists between files' position and the detection of the \cornercase.
However, in the case of \missingbreak our test achieved  ${\chi}^2 (df=1, N =106)=0.88)$, p-value = $0.346$. Therefore, for this defect we could not draw any conclusion on the fact that its position influenced developers' ability to find it. 

For the \cornercase, we also computed the phi-coefficient to measure the strength of the association obtaining a value of 0.247, which is close to a moderate positive correlation~\cite{glen2016}.

\begin{table}[ht]
	\centering 
	\caption{Logistic regression models for RQ2.} \label{tab:rq2Logistic}
	{\rowcolors{3}{gray05}{white}
	\begin{tabular}{lrcl|rcl}
		& \multicolumn{3}{c|}{\cellcolor[HTML]{C0C0C0}\textbf{Dep. Var. = CC Found}} & \multicolumn{3}{c}{\cellcolor[HTML]{C0C0C0}\textbf{Dep. Var. = MB Found}} \\
		&  \multicolumn{1}{c}{\cellcolor[HTML]{E8E2E1}\textbf{Estim.}} & \multicolumn{1}{c}{\cellcolor[HTML]{E8E2E1}\textbf{S.E.}} & \multicolumn{1}{c|}{\cellcolor[HTML]{E8E2E1}\textbf{Sig.}} &
		 \multicolumn{1}{c}{\cellcolor[HTML]{E8E2E1}\textbf{Estim.}} & \multicolumn{1}{c}{\cellcolor[HTML]{E8E2E1}\textbf{S.E.}} & 
		\multicolumn{1}{c}{\cellcolor[HTML]{E8E2E1}\textbf{Sig.}}	\\
		\hline
		Intercept & $-1.113$ &1.570 &  &$ -1.573$ &1.536 & \\
		Position & $-0.960$ &0.409 & * & 0.388 & 0.418 & \\
		CRDur. & $-3\,10^{-4}$ &0.009 & & 0.005 & 0.010 & \\
		DevExp & $-0.024$ &0.225 & & 0.052 &0.227 & \\
		JavaExp & 0.120 & 0.189 & & 0.251 & 0.191 & \\
		CRExp & 0.045 & 0.228 & & 0.326 & 0.236 & \\
		OftenProg & 0.171  & 0.323 & & 0.130 & 0.312 & \\
		OftenCR & 0.024 &0.220 & & $-0.547$ & 0.235 & * \\
		Interrupt. & 0.005 & 0.156 & &$-0.056$ & 0.158 & \\
		\hline
		& \multicolumn{6}{c}{Sig. codes: * p < 0.05} \\
	\end{tabular}
	}
\end{table}
 
 To investigate the association between the detection of a defect, its position in the code change, and other possible confounding factors (reported in \Cref{tab:logVariables}), we built two logistic regression models, whose results are shown in \Cref{tab:rq2Logistic}. These findings are aligned with the results of the \textit{Chi-square} test: The position of the \cornercase is statistically significant correlated with developers' ability to find it (p-value $< 0.05$). 
 
 \roundedbox{The relative position of the file containing the \cornercase influences the likelihood of participants in detecting the defect. Participants are less likely to find the defect when its file is presented last. This effect is not significant for the \missingbreak.}
 
At the end of the code review task we asked participants if they thought the position of the defect had an influence on their ability to find it. The vast majority of them replied negatively ($72.6\%$ for the \cornercase and $76.4\%$ for the \missingbreak). However, our results showed how, on the contrary, the position of the bug had an effect on participants' ability to find it. This shows a mismatch between what developers \textit{think} affects the code review outcome and which factors \textit{actually} play a role in it. 
 
\medskip
\noindent\textbf{4.2.2 Visible Files}
	
\begin{figure}[t]
 	\centering
 	\includegraphics[width=0.5\textwidth]{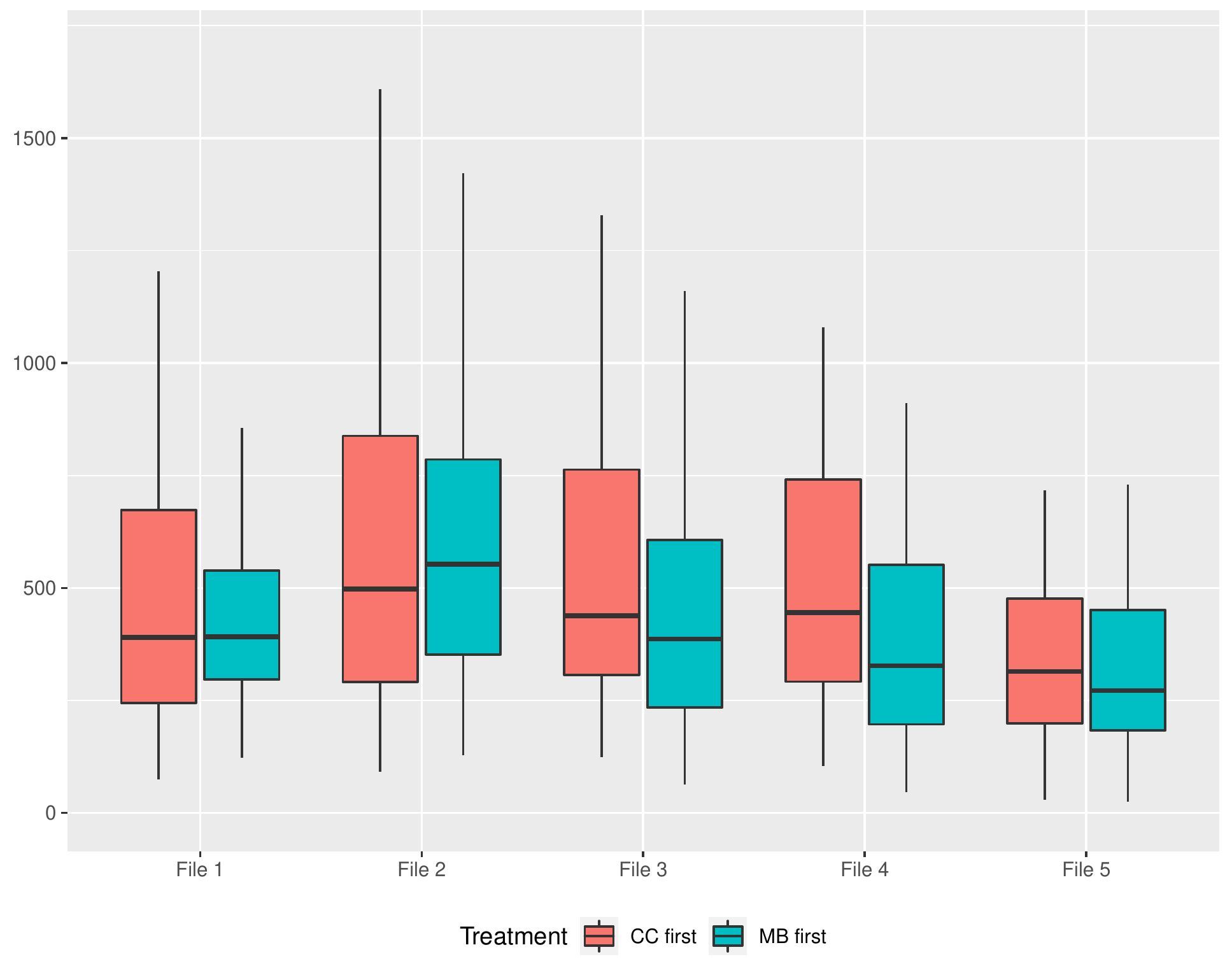}
 	\caption{Time (in seconds) participants visualized each file. To improve the clarity, we limited the size of the Y axis.} 
 	\label{fig:summaryTime}
 \end{figure}

\noindent In the experiment, we measured the time that each file was visible on the participants' screen as a way to investigate further the effect of files' position on code review. \Cref{fig:summaryTime} reports the time (in seconds) that participants in the two treatments spent with each file visible on the screen. We removed from this analysis participants who declared to have been interrupted for more than ten minutes during the review task, as their data may bias this analysis. 
 
 On average, participants regardless of their treatment (\tOne or \tTwo) spent more time looking at the first file compared to the last one. On average, participants spent 8.58 minutes with the first file displayed as opposed to 6.09 for the last file. 
 
 
 To verify whether the difference between the time participants visualized the first and last file is statistically significant, we performed a \textit{Mann–Whitney U test}. For the whole participants (without dividing them by the treatment), our test obtained a p-value of 0.002, therefore showing that the two variables are indeed different. Dividing the participants by treatment, our test achieved a p-value of 0.022 for the treatment \tOne, and 0.031 for treatment \tTwo. These results confirm those for the general population. 
 
\medskip
\roundedbox{During the review task, participants spent significantly more time with the first file being visible compared to the last file. }

\noindent\textbf{4.2.3 Robustness Testing}
	
\noindent To challenge the validity of our findings and strengthen their goodness, we employed \textit{robustness testing}~\cite{neumayer2017robustness}. 

\smallskip
\noindent
\textbf{Participants' groups are not homogeneous.} Participants' experience in programming and reviewing might impact their performance during code review and, therefore, influence the results. To verify that the participants assigned to the two treatments have similar characteristics, we performed a \textit{Chi-square test of homogeneity} on the variables measuring participants' experience and practice (\eg participants' experience with Java). We obtained p-values above 0.1, therefore, not revealing significant differences between the two samples of participants. Moreover, we compared the time spent on the code review task of participants belonging to the two treatments. To this aim, we performed a \textit{Mann-Whitney U test} (p-value of 0.453), which did not detect any difference between the two groups. 

\smallskip
\noindent
\textbf{One defect might influence participants in finding the other.} Finding one defect might have biased participants towards finding the other one. For instance, participants who found one bug might have stopped looking for other defects assuming they found the only issue in the code. Another possibility is that one defect can give unintentional indications to participants to identify the other one.
However, the results reported in \Cref{tab:bugsFoundTotal} counter the existence of such bias: the number of participants who found only one bug is similar to the one of those who found both bugs. 
We further performed a \textit{Chi-square} test to verify if finding one defect led participants to also find the second one. Our test obtained a p-value of 0.888, thus suggesting this bias not to affect the experiment. 

\smallskip
\noindent
\textbf{The defects are too easy/difficult.} The selected defects might have been too easy, or on the contrary, too hard to identify for the participants. To mitigate this possible bias we (1) selected defects among common Java mistakes based on relevant literature~\cite{switch1, Spadini2019, jeng1994simplified} and discussed them among the authors, and (2) verified our choice through a pilot study. Despite these measures, participants in the experiment might have still faced difficulties finding these defects (or identified them too easily). However, the results reported in \Cref{tab:bugsFoundTotal} show that the number of participants who found no defects, only one defect, or both defects is almost evenly distributed, showing that no defect was too trivial or hard to find.

\smallskip
\noindent
\textbf{A low number of participants.}
A number of participants too low might bias the significance of our findings. The power analysis we run (\Cref{sec:met:rq2}) showed that we needed at least 92 participants. Our experiment exceeded this number: It was conducted by a total of \validParticipants valid participants.

\section{Threats to Validity}

\textbf{Construct validity.} The code changes selected for the review task might have introduced bias in the experiment. To mitigate this threat, the first and second authors prepared the code changes and selected the defects contained in them. The other authors also checked the goodness of the code review task. 

The scenario represented in our experiment might differ from a real-world review. To reduce this threat, we adopted the following measures: we (1) created a code change as similar as possible to real ones (\eg including documentation), (2) seeded defects that commonly happen in real-world scenarios~\cite{spadini2020primers, switch1}, and (3) used a review interface similar to widely used code review tools.

\smallskip
\noindent
\textbf{Internal validity.} During code review, developers, instead of writing multiple similar comments, might leave only one comment requiring changes in all similar subsequent instances of an issue. Such comments might introduce bias in the results of our investigation in RQ$_1$. For this reason, one of the authors manually inspected a randomly selected subset of 100 PRs, looking for such comments. Only eight PRs contained instances of these comments. Considering this proportion, we computed a statistically significant sample size with confidence level $95\%$ and margin of error $5\%$, obtaining a sample size of 54, which confirmed that 100 PRs are sufficient to establish the proportion of this factor in our results. 

In RQ$_2$, before analyzing the results, we manually inspected the participants' logs. We removed all participants that did not take the review task seriously: Participants who spent less than ten minutes on the review and left no remarks. 
Since our experiment relied on an online platform, the participants might have completed the review task with different set-ups and in different environments. However, this reflects the various conditions in which developers work in real-world scenarios. To mitigate the threats that interruptions might constitute for the validity of our study, we asked participants to state how long they were interrupted during the code review task and included their answers in our statistical analyses. 


\smallskip
\noindent
\textbf{External validity.} In RQ$_1$, the selection of the projects might have introduced bias in our investigation. To reduce potential bias caused by the choice of a specific project, we considered a vast set of projects from different domains, all with star-count above 1,000. However, we cannot exclude the possibility that the choice of different projects might lead to different results. Further studies are needed to verify the generalizability of our findings. 

Participants in our experiment possess a vast set of different backgrounds, yet our sample is not representative of all developers. 
Moreover, the specific defects we chose might have influenced the experiment results. Although these defects have been reported in the literature as common defects~\cite{jeng1994simplified, spadini2020primers, switch1}, they have specific characteristics that do not generalize to all other types of defects.

Furthermore, the size of the code change contained in the review task might have influenced the observed results. We carefully chose the review code change size to be a trade-off between a too small review (where the effect of the files' positions is likely not to matter) and a too-large task, which would have resulted in a higher participants drop-out rate. Although the results of RQ$_1$ suggest that this effect should be present also in the code change with a higher number of files, further studies are needed to corroborate our findings in different scenarios. 

\bluetext{In our experiment, we considered only code written in Java. This was done to keep the necessary number of participants (\Cref{sec:met:rq2}) within an achievable range.
In fact, considering multiple programming languages would have required to significantly extend the number of participants due to the following reasons.
First, although languages such as Java, C\#, and Python can be used to write code in a similar way, in reality they tend to follow different coding conventions and idioms~\cite{alexandru2018usage, phan2020teddy, pythonicMeans}. Therefore, using multiple languages poses the issue of either not following the language's idioms (thus making the code snippet less ``natural'') or having snippets substantially different across languages. Second, past research has provided evidence on different programmers' behavior when comprehending code written in different languages~\cite{turner2014eye}. Third, work on programming languages learning shows that novices face different difficulty levels with different programming languages~\cite{mannila2006simple}. This might be caused by the different cognitive load that each language poses. 
For these reasons, we cannot exclude that different programming languages might lead to different results.
Further studies can be devised and conducted to investigate whether and how the results of our experiment change considering different programming languages.
}

\section{Discussion}

In this section, we discuss the main implications of our findings.

\smallskip
\noindent
\textbf{The importance of being first.} Our investigation provides evidence that the position of a file influences developers' review effectiveness: Developers have 64\% lower odds to find a defect when it is in the file shown last. 
Currently, code review tools (\eg GitHub or Phabricator) display files in alphabetical order. Our results suggest that more principled ways of presenting the files in a code review can be used to support reviewers' effectiveness. 

For example, tools could display first those files that are more critical. 
Previous work focused on identifying salient classes in a code change (\ie classes that subsequently cause the modification of other classes in the change-set)~\cite{huang2018salient}. Such classes can be used as the point from which developers should start a revision. 
\bluetext{On the contrary, another possibility would be to focus on identifying \textit{arid files} and placing them last in a code review, taking inspiration from the approach developed at Google to identify \textit{arid lines}~\cite{petrovic2018state}. Such an approach would allow developers to prioritize the review of more relevant files.}


Other studies proposed an ordering approach to re-organize files in a code change to support reviewers' preferences~\cite{Baum2017, Baum2019}. This approach focuses on how files to review should be grouped, suggesting to show files sequentially sharing a link (\eg method call - method declaration). 
However, this ordering theory does not consider the absolute position of the files (\ie which ones should be shown first). The ordering theory can be expanded to take into account our findings.
\bluetext{Moreover, previous work~\cite{Baum2017} reported how developers adopt different tactics when reviewing code (\eg starting from newly added files or files perceived as easier to review). For this reason, future code review tools might include a way to let reviewers customize the order of the files to review to fit different reviewers' needs and counter the effect of the relative position of the files. 
}

\bluetext{Finally, if changing the alphabetical order of the files in a code review tool is too unnatural for reviewers, one could consider using warnings to point reviewers towards the file from which they should start the review. To this aim, new solutions could be devised taking inspiration from previous studies on program-analysis tools~\cite{johnson2013don, henley2018cfar}, where the authors investigated how warnings should be displayed to be welcomed by developers. These warnings could, for instance, make developers aware of biases during code review (\eg as created by the relative file position)}. Future studies can be designed and carried out to compare the effectiveness of using warnings as opposed to changing the position of files in a code review tool.

\smallskip
\noindent
\textbf{Not all defects are the same.} In our experiment, participants' effectiveness in detecting the \cornercase was influenced by the relative position of the defective file. However, this was not the case for the \missingbreak.

If we consider the nature of these bugs, we notice that they may require a different effort to be found, even though they were found with the same probability overall (as reported in \Cref{tab:bugsFound}).
On the one hand, the \cornercase requires the reviewer to read and understand the documentation, then comprehend the code and identify a mismatch in the corresponding implementation. As one participant explained when asked why they did not find the \cornercase: ``I didn't read the JavaDoc carefully enough, and there was not anything obviously wrong with the code.''
On the other hand, the \missingbreak can be recognized as a pattern, even if one does not fully understand the semantic of the code. One participant who found it explained: ``I am accustomed to looking for break statements on every switch expression because I have missed them in my own code in the past.''

Therefore, the underlying reason for the difference in files' position effect could be that these two bugs require a different cognitive effort to be found. These bugs could impose a different load on participants' working memory, making, in turn, the effect of the defect position more or less prominent. We did not investigate this further, but studies can investigate this hypothesis and determine the role of cognitive effort in defect finding during review.


\smallskip
\noindent
\textbf{The hidden psychological factors of code review.} At the end of our experiment, we showed the participants the defects we seeded in the code. When asked whether the position of the bugs might have had an impact on their ability to detect them, a vast amount of the developers replied negatively. A total of 77 ($72.6\%$) participants rejected the idea that files' position influenced their ability to find for the \cornercase. 
However, the results of our experiment found evidence of the contrary. The impact of file position on reviewers was also corroborated by the first part of our study when associating file position and review activity.

This mismatch between participants' perceptions and actual results contributes to the discourse around the importance and the impact of cognitive biases that may affect developers during different software engineering activities~\cite{mohanani2018cognitive}.
After all, code review is a human effort and as humans we are influenced by biases and ``\textit{hidden}'' psychological factors of which we are not naturally aware.



\smallskip
\noindent
\textbf{Beyond code review.} Our results provided initial evidence on the effect of files' position on developers' activity and effectiveness in code review. However, this phenomenon might also affect other aspects of software engineering. 

Previous studies investigated factors affecting developers' adoption of program-analysis tools, showing that the way in which warnings are displayed is critical~\cite{johnson2013don, henley2018cfar}. Among the many factors identified to improve how warnings are displayed (\eg warning should be well motivated~\cite{johnson2016cross}), our results may indicate that the position of the warning can play a role. Similarly to files, the position of the warning might affect, for instance, the attention that developers pay to them. 
Future studies can evaluate whether the effect of warnings' position influences developers. This effect might be particularly significant, for instance, for those complex warnings that require developers' effort to be understood and solved.

\section{Conclusion}

In this study, we investigated whether the relative position of a file has an impact on code review. To do so, we devised a two-step investigation to collect complementary evidence: We (1) collected and analyzed data from \numberPRs Pull Requests (PRs) belonging to \numberProjects open-source Java projects and (2) conducted an online controlled experiment with \validParticipants participants. 

In the first step of our investigation, we focused on reviewers' activity, investigating the relationship between the relative position of file in a PR and the number of comments it receives.
Having found evidence of a significant correlation between file position and review activity, we moved to the second step of our study to collect different evidence. We devised an online experiment where participants had to perform review a code change in which we two seeded defects (Corner Case and Missing Break). Participants were randomly assigned to one of two possible treatments: \textit{\tOne} (where the file with the \cornercase was shown first, while the one with the \missingbreak was shown last) and \textit{\tTwo} (where the order of the files was reversed). 

Our experiment found that the position of the file containing the \cornercase influences the likelihood of finding this defect. Specifically, 34 participants found the \cornercase when in the first file, while only 18 found it in the last. This effect was not significant for the less cognitive demanding \missingbreak.
Overall, our study provides evidence that the position in which files are presented during code review has an impact on code review's outcome. This finding has implications for code review practices and review tool design, and may suggest that a similar effect may be present in other software engineering contexts.


\begin{acks}
The authors would like to thank the anonymous reviewers for their thoughtful and important comments, which helped improving our paper. Fregnan and Bacchelli gratefully acknowledge the support of the Swiss National Science Foundation through the SNSF Projects No. PP00P2\_170529 and 200021\_197227. D'Ambros gratefully acknowledges the financial support of the Swiss National Science Foundation through the NRP-77 project 187353.
\end{acks}

\bibliographystyle{ACM-Reference-Format}
\bibliography{main-firstposition}


\end{document}